\documentclass[11pt]{article}
\usepackage{amsmath}
\numberwithin{equation}{section}

\begin{document}

\def\been{\begin{enumerate}}
\def\enen{\end{enumerate}}
\def\beit{\begin{itemize}}
\def\enit{\end{itemize}}
\def\be{\begin{equation}}
\def\en{\end{equation}}
\def\bear{\begin{eqnarray}}
\def\enar{\end{eqnarray}}
\def\beas{\begin{eqnarray*}}
\def\enas{\end{eqnarray*}}
\def\bera{\begin{eqnarray}}
\def\enra{\end{eqnarray}}

\renewcommand{\theequation}{\thesection.\arabic{equation}}

\newcommand{\gsim}{\mbox{\raisebox{-1.ex}{$\stackrel
        {\textstyle>}{\textstyle\sim}$}}}
\newcommand{\lsim}{\mbox{\raisebox{-1.ex}{$\stackrel
        {\textstyle<}{\textstyle \sim}$}}}
\newcommand{\qed}{\hbox{\rule[-2pt]{6pt}{6pt}}}

\def\ptl{\partial}
\newcommand{\half}{\frac{1}{2}}
\def\CC{{\cal C}}
\def\buildchar#1#2#3{\null \! \mathop {\vphantom {#1}
\smash #1}\limits ^{#2}_{#3}\!\null }
\def\ut#1{\buildchar{#1}{}{^\sim}\/}
\def\dtri{\tilde{E}}
\def\CH{{\cal C_H}}
\def\CM{{\cal C_M}}
\def\CG{{\cal C_G}}
\def\mm[#1,#2,#3,#4]
 {\left(\matrix{#1 & #2 \cr #3& #4}\right)}
\def\largefbox#1#2{\begin{center}
                         \setlength{\fboxsep}{0.25cm}\fbox{\parbox[c]{#1}{#2}}
                    \end{center}}
\def\Largefbox#1#2{\begin{center}
                         \setlength{\fboxsep}{0.5cm}\fbox{\parbox[c]{#1}{#2}}
                    \end{center}}
\def\boxwidth{160mm}  
\def\boxwidthitem{150mm}  
\newenvironment{indention}[1]{\par
\addtolength{\leftskip}{#1}
\begingroup}{\endgroup\par}


\title{{\bf Kinematic self-similar solutions in general relativity}}

\author{Hideki Maeda  \\ {\tt hideki@gravity.phys.waseda.ac.jp}  \\
      Advanced Research Institute for Science and Engineering, \\
Waseda University, Okubo 3-4-1, Shinjuku, Tokyo 169-8555, Japan \\
      ~\\ 
      Tomohiro Harada \\  {\tt T.Harada@qmul.ac.uk}  \\
      Astronomy Unit, School of Mathematical Sciences, \\
Queen Mary, University of London,
Mile End Road, London E1 4NS, UK
      }

\date{\today} 
\maketitle

\begin{abstract}
The gravitational interaction is scale-free in both Newtonian gravity 
and general theory of relativity.
The concept of self-similarity arises from this nature.
Self-similar solutions reproduce themselves as the scale changes.
This property results in great simplification of the governing 
partial differential equations.
In addition, some self-similar solutions can describe the asymptotic
behaviors of more general solutions.
Newtonian gravity contains only one dimensional constant, the 
gravitational constant, while the general relativity 
contains another dimensional constant, the speed of light, besides
the gravitational constant.
Due to this crucial difference, incomplete similarity 
can be more interesting in general relativity than in Newtonian gravity.
Kinematic self-similarity has been defined and studied
as an example of incomplete similarity 
in general relativity, in an effort to
pursue a wider application of self-similarity in general relativity.
We review the mathematical and physical aspects of 
kinematic self-similar solutions in general relativity.
\end{abstract}

\newpage
\setcounter{tocdepth}{2}
\tableofcontents
\newpage

\section{Introduction}
Scale-invariance is one of the most fundamental characteristics 
of gravitational interaction in both Newtonian gravity and general relativity.
This implies that if we consider appropriate matter fields,
the governing partial differential equations are invariant 
under scale transformation. Due to this feature of the governing
equations, there are self-similar solutions, which are 
invariant under the scale transformation. 
Self-similarity assumption enables us to simplify the 
governing equations. Self-similar solutions 
have a wide range of applications in astrophysics.
See~\cite{cc2000b} for a recent review of self-similar solutions
in general relativity.
See~\cite{barenblatt1996} for self-similarity in more general contexts.

When a theory has no characteristic scale, we can expect 
scale-invariance of the theory.
In Newtonian gravity, the gravitational constant $G$, with dimension 
$M^{-1}L^3T^{-2}$, is the only dimensional physical constant in the field
equations, where $M$, $L$ and $T$ denote the dimensions of 
mass, length and time, respectively. 
It is impossible to construct a physical scale 
only from $G$. In general relativity, there exists another physical
constant $c$, which is the speed of light, with dimension
$LT^{-1}$. In spite of these two dimensional constants, no 
characteristic length scale can be constructed from these 
physical constants. However, due to the existence of these two dimensional 
constants, general relativity is qualitatively different from Newtonian
gravity with respect to scale invariance. 
If we consider quantum gravity, 
the Planck constant $h$ appears,
with dimension $ML^2T^{-1}$, so that there exists a characteristic
scale $l_{\rm pl} \equiv G^{1/2}h^{1/2}/c^{3/2}$, which is 
called the Planck length. Therefore, in the quantum theory of gravity,
it is plausible that the scale invariance of the theory 
is broken down. Hereafter in this review we focus on Newtonian gravity 
and general relativity.
We follow the sign conventions of~\cite{gravitation}
for the metric, Riemann and Einstein tensors.

\section{Self-similarity in Newtonian gravity}
Since Newtonian gravity postulates an absolute system of space and time,
we can directly apply the general formulation of
self-similarity to this system~\cite{barenblatt1996}.
A solution is called self-similar, if a dimensionless quantity 
$Z(t,{\vec x})$  made of the solution is of the form 
\begin{eqnarray}
Z(t,{\vec x})=Z\left(\frac{{\vec x}}{a(t)}\right),
\end{eqnarray}
where ${\vec x}$ and $t$ are independent space and time coordinates,
respectively, and $a(t)$ is a function of $t$. This implies that
the spatial distribution of the characteristics of motion remains
similar to itself at all times during the motion. If the function $a(t)$
is derived from dimensional considerations alone, i.e., if it is
uniquely determined so that ${\vec x}/a(t)$ is dimensionless,
the self-similarity is called complete similarity or
similarity of the first kind~\cite{barenblatt1996}. 
In more general situations, the characteristic length 
or time scale may be constructed by the dimensional constants 
in the system. Then, the function $a(t)$ cannot be uniquely 
determined from dimensional considerations alone.
In such cases, self-similarity is called incomplete similarity
or similarity of the second kind~\cite{barenblatt1996}. 
For example, when we have the constant sound speed $c_{\rm s}$
and no characteristic scale, then $a(t)$ is uniquely determined
as $a(t)=c_{\rm s}t$. In this case, the similarity is 
called complete. However, when we have a characteristic 
length scale $l$ besides the sound speed $c_{\rm s}$, 
then $a(t)=l^{1-\alpha}(c_{\rm s}t)^{\alpha}$
is possible and the constant $\alpha$ may not be determined
from the governing equations. 
In this case, the similarity is called incomplete.
The constant $\alpha$ may be determined by 
boundary conditions. It should be noted that the dimensional 
constant could appear not only from governing equations 
but also from boundary conditions.

Here, we give two important examples of completely self-similar solutions 
in Newtonian self-gravitating fluid mechanics. The basic field equations 
for spherically symmetric hydrodynamics of a self-gravitating 
ideal gas in Eulerian description are given by
\begin{eqnarray}
&&\frac{\partial\rho}{\partial t}+\frac{1}{r^2}\frac{\partial}{\partial r}(r^2\rho v)=0, \label{continuity}\\
&&\frac{\partial}{\partial t}(\rho v)+\frac{1}{r^2}\frac{\partial}{\partial r}(r^2\rho v^2)+\frac{\partial p}{\partial r}+\rho\frac{G M}{r^2}=0,  
\label{euler}\\
&&\frac{\partial M}{\partial t}+v\frac{\partial M}{\partial r}=0,  
\label{masseq1}\\
&&\frac{\partial M}{\partial r}=4\pi r^2 \rho, \label{masseq2}
\end{eqnarray}
where $\rho,v,M$ and $G$ denote the mass density, radial velocity, total mass inside the radial coordinate $r$, and gravitational constant, respectively.

\subsection{Isothermal gas}
First we consider an isothermal gas as a gravitational
source. Since the isothermal gas is a relevant description of 
cold molecular clouds in galaxies, self-similar solutions have 
been intensively studied in Newtonian 
gravity in modeling the star formation 
process~\cite{larson1969,penston1969,shu1977,hunter1977,ws1985}.
It has been revealed that self-similar solutions play 
important roles in the gravitational collapse of an isothermal 
gas~\cite{fc1993,ti1999,hms2003}.
The stability of these self-similar solutions have been 
studied~\cite{op1988b,hn1997,hm2000a,hm2000b}.
A new insight has been obtained in this system 
in the context of critical behavior in 
gravitational collapse~\cite{mh2001,hms2003}.

For an isothermal gas that obeys $p=c_{\rm s}^2 \rho$, where $c_{\rm s}$ is
the constant speed of sound with dimension $LT^{-1}$, 
it is impossible to construct a characteristic scale from $c_{\rm s}$ and $G$. 
We introduce the dimensionless self-similar
coordinate 
\begin{equation}
z=\frac{c_{\rm s} t}{r},
\end{equation}
for self-similar solutions. 
Then we also introduce the dimensionless functions $U$, $P$ and $m$:
\begin{eqnarray}
v(r,t)&=&-c_{\rm s} U(r,t), \\
\rho(r,t)&=&\frac{c_{\rm s}^2 P(r,t)}{4\pi G r^2} , \\
\quad M(r,t)&=&\frac{c_{\rm s}^3 t m(r,t)}{G}.
\end{eqnarray}
We assume that the above-defined functions $U$, $P$ and $m$ depend only
on $z$.
From this assumption, equations (\ref{continuity})--(\ref{masseq2}) become 
\begin{eqnarray}
U'&=&\frac{(zU+1)[P(zU+1)-2]}{(zU+1)^2-z^2}, \\
P'&=&\frac{zP[2-P(zU+1)]}{(zU+1)^2-z^2}, \\
m&=&P(U+1/z),\\
-z^2m'&=&P,
\end{eqnarray}
where the prime denotes the derivation with respect to $z$. The
self-similar solutions for an isothermal gas are obtained from these
ordinary differential equations. Self-similar solutions scale for
the scale transformations ${\bar t}= at,~{\bar r}= ar$ as 
\begin{eqnarray}
v({\bar r},{\bar t})&=&v(r,t), \\
\rho({\bar r},{\bar t})&=&\frac{\rho(r,t)}{a^{2}}, \\
M({\bar r},{\bar t})&=&aM(r,t),
\end{eqnarray}
where $a$ is a constant.
The basic equations for self-similar solutions are singular at the
center and at the point at which $(zU+1)^2-z^2=0$ is satisfied, which is
called a sonic point.

\subsection{Polytropic gas}
Next, we consider a polytropic gas as a gravitational
source. A polytropic gas obeys the equation of state $p=K\rho^{\gamma}$, where
$\gamma$ is the dimensionless constant called 
the adiabatic exponent and $K$ is a constant with dimension
$M^{1-\gamma}L^{3\gamma-1}T^{-2}$. As in the isothermal gas system, it is impossible to construct 
a characteristic scale only from $G$ and $K$ if $\gamma\ne 2$.
For the exceptional case, $\gamma=2$, the system has a 
characteristic length scale $l=\sqrt{K/G}$ but even in this case
the self-similar variable $z$ is uniquely constructed.
Then, complete similarity is applicable to this system.
Self-similar solutions in this system have been 
studied~\cite{yahil1983,ss1988}.
The stability of these solutions have been 
studied~\cite{hm2000a,hm2000b}.

For the polytropic case, we introduce the dimensionless self-similar
coordinate 
\begin{equation}
z=\frac{\sqrt{K}(-t)^{2-\gamma}}{(4\pi G)^{(\gamma-1)/2}r}.
\end{equation} 
Then we also introduce the dimensionless functions $U$, $P$ and $m$:
\begin{eqnarray}
v(r,t)&=&-(4\pi G)^{(1-\gamma)/2}\sqrt{K}(-t)^{1-\gamma}U(r,t),\\
\rho(r,t)&=&\frac{K^{1/(2-\gamma)} P(r,t)}{(4\pi G)^{1/(2-\gamma)} r^{2/(2-\gamma)}} , \\
M(r,t)&=&\frac{K^{3/2}(-t)^{4-3\gamma} m(r,t)}{(4\pi)^{3(\gamma-1)/2}G^{(3\gamma-1)/2}}.
\end{eqnarray}
We assume that the above-defined functions $U$, $P$ and $m$ depend 
only on $z$.
In the polytropic case, the sonic point is defined by
$(2-\gamma-zU)^2-\gamma z^{2/(2-\gamma)}=0$. Self-similar solutions
scale 
for the scale transformations ${\bar t} =at,~{\bar r}= a^{2-\gamma}r$, as 
\begin{eqnarray}
v({\bar r},{\bar t})&=&a^{1-\gamma}v(r,t), \\ 
\rho({\bar r},{\bar t})&=&\frac{\rho(r,t)}{a^{2/(2-\gamma)}}, \\ 
M({\bar r},{\bar t})&=&a^{4-3\gamma}M(r,t),
\end{eqnarray}
where $a$ is a constant.
In this case, the scaling rates for $r$ and $t$, which keep $z$ constant, 
are different from each other. 

It should be again emphasized that in both the isothermal and polytropic
cases, the self-similarity is complete since the self-similar variable
$z$ can be obtained from dimensional considerations alone. This is 
because there are only two dimensional constants in the system, 
while there are three independent dimensions $M$, $L$ and $T$.

\section{Self-similarity in general relativity}
\subsection{Homothety}
In general relativity, the concept of self-similarity is not so
straightforward because general relativity 
has general covariance against coordinate 
transformation. This implies that the definition should be made 
covariantly in general relativity. In the following,
we use units where the speed of light $c$ is unity. In this choice
of units, $T=L$ is obtained and the velocity is dimensionless.

In general relativity, the term self-similarity can be used in two
ways. 
One is for the properties of spacetimes, the other is for the  
properties of matter fields. These are not equivalent in general. The
self-similarity in general relativity was defined for the first time by
Cahill and Taub~\cite{ct1971}.  Self-similarity is defined 
by the existence of a homothetic  
vector ${\mbox{\boldmath $\xi$}}$ in the spacetime, which satisfies
\begin{eqnarray}
{\cal  L}_{\mbox{\boldmath $\xi$}} g_{\mu\nu} = 2\alpha g_{\mu\nu} \label{gss1},
\end{eqnarray}
where $g_{\mu\nu}$ is the metric tensor, 
${\cal L}_{\mbox{\boldmath $\xi$}}$ denotes Lie differentiation
along ${\mbox{\boldmath $\xi$}}$ and $\alpha$ is a constant~\cite{ct1971}. 
This is a special type of conformal Killing vectors.
This self-similarity is called homothety. 
If $\alpha \ne 0$, then it can be set to be unity by a 
constant rescaling of ${\mbox{\boldmath $\xi$}}$. 
If $\alpha=0$, i.e. ${\cal L}_{\mbox{\boldmath $\xi$}} g_{\mu\nu}=0$, then
${\mbox{\boldmath $\xi$}}$ is a Killing vector. 

Homothety is a purely geometric property
of spacetime so that the physical quantity does not necessarily exhibit 
self-similarity such as ${\cal  L}_{\mbox{\boldmath $\xi$}} Z = d Z$, where $d$ is a
constant and $Z$ is, for
example, the pressure, the energy density and so on.
From equation (\ref{gss1}) it follows that
\begin{eqnarray}
{\cal  L}_{\mbox{\boldmath $\xi$}} R^{\mu}\,_{\nu\sigma\rho} =0,\label{curvecoll}
\end{eqnarray}
and hence 
\begin{eqnarray}
{\cal  L}_{\mbox{\boldmath $\xi$}} R_{\mu\nu} &=&0 \label{riccicoll},\\
{\cal  L}_{\mbox{\boldmath $\xi$}} G_{\mu\nu} &=&0 \label{mattercoll}.
\end{eqnarray}
A vector field ${\mbox{\boldmath $\xi$}}$ that satisfies equations (\ref{curvecoll}),
(\ref{riccicoll}) and (\ref{mattercoll}) is called a curvature
collineation, a Ricci collineation and a matter collineation,
respectively. 
It is noted that equations~(\ref{curvecoll}), 
(\ref{riccicoll}) and (\ref{mattercoll})
do not necessarily mean that ${\mbox{\boldmath $\xi$}}$
is a homothetic vector.
We consider the Einstein equations
\begin{eqnarray}
G_{\mu\nu}=8\pi GT_{\mu\nu} \label{einstein},
\end{eqnarray}
where $T_{\mu\nu}$ is the energy-momentum tensor. 
If the spacetime is homothetic,
the energy-momentum tensor of the matter fields 
must satisfy
\begin{eqnarray}
{\cal  L}_{\mbox{\boldmath $\xi$}} T_{\mu\nu} =0,\label{emcoll}
\end{eqnarray}
through equations~(\ref{einstein}) and (\ref{mattercoll}).
For a perfect fluid case, the energy-momentum tensor takes the form of
\begin{eqnarray}
T_{\mu\nu} = (p+\mu) u_{\mu}u_{\nu}+ pg_{\mu\nu},
\label{perfect_fluid}
\end{eqnarray}
where $p$ and $\mu$ are the pressure and the energy density,
respectively. Then,
equations~(\ref{gss1}) and (\ref{emcoll}) result in
\begin{eqnarray}
{\cal  L}_{\mbox{\boldmath $\xi$}} u^{\mu} &=&-\alpha u^{\mu},\\
{\cal  L}_{\mbox{\boldmath $\xi$}} \mu &=&-2\alpha \mu, \label{ssp}\\
{\cal  L}_{\mbox{\boldmath $\xi$}} p &=&-2\alpha p. \label{ssmu}
\end{eqnarray}
As shown above, for a perfect fluid, the self-similarity of the spacetime and that of the
physical quantity coincide. However, this fact does not
necessarily hold for more general matter fields. 

For spherically symmetric homothetic spacetimes, 
we can assume that there is a coordinate
system $t$ and $r$ such that all dimensionless variables are 
functions of a single dimensionless self-similar variable $\xi \equiv
r/t$. The solution is invariant under scale transformation 
$\overline{t} = at$, $\overline{r} = ar$ for any constant $a$. 
Thus the self-similar variables can be determined from dimensional
considerations in the case of homothety. Therefore, we can conclude 
homothety as the general relativistic analogue of complete similarity.  

From the constraints (\ref{ssp}) and (\ref{ssmu}), we can show that if
we consider the barotropic equation of state, i.e.,
$p=f(\mu)$, then the equation of state must have the form $p=K\mu$,
where $K$ is a constant. 
This class of equations of state contains a dust fluid ($K=0$),
a radiation fluid ($K=1/3$) and a stiff fluid ($K=1$)
as special cases.
Other important matter fields that are compatible with homothety
are a massless scalar field and a scalar field with an exponential potential.

\subsection{Kinematic self-similarity}
Although homothetic solutions can contain several interesting
matter fields, the matter fields compatible with 
homothety are rather limited.
In more general situations, matter fields will have 
intrinsic dimensional constants.
For example, when we consider a polytropic equation of state, 
such as $p=K\mu^{\gamma}$, the constant $K$ has dimension 
$M^{1-\gamma}L^{3(\gamma-1)}$, where we should be reminded
that we have chosen the light speed $c$ to be unity.
We can also consider a massive scalar field, where the mass
of the scalar field has dimension $M$.
In such cases, it is impossible to assume homothety because the 
system has a characteristic scale.
By analogy, we can consider the general relativistic counterpart
of incomplete similarity.
From comparison with self-similarity for a polytropic gas 
in Newtonian gravity,
kinematic self-similarity has been defined in the context
of relativistic fluid mechanics as an example of 
incomplete similarity~\cite{ch1989,ch1991,coley1997}.
It should be noted that the introduction of incomplete 
similarity to general relativity is not unique.
For example, partial self-similarity has been defined and 
applied to inhomogeneous cosmological 
solutions~\cite{tomita1981a,tomita1981b,tomita1997c}.

A spacetime is said to be kinematic self-similar if it admits 
a kinematic self-similar vector ${\mbox{\boldmath $\xi$}}$ which satisfies the conditions
\begin{eqnarray}
{\cal{L}}_{\mbox{\boldmath $\xi$}} h_{\mu\nu} &=&2\delta h_{\mu\nu},\label{kss}\\
{\cal{L}}_{\mbox{\boldmath $\xi$}} u_{\mu} &=&\alpha u_{\mu},\label{gss}
\end{eqnarray}
where $u^{\mu}$ is the four-velocity of the fluid and 
$h_{\mu\nu} =g_{\mu\nu}+u_{\mu}u_{\nu}$ is the projection tensor,
and $\alpha$ and $\delta$ are
constants~\cite{ch1989,ch1991,coley1997}. 
If $\delta\ne 0$, the similarity transformation
is characterized by the scale-independent ratio 
$\alpha/\delta$, which is referred to as the similarity index. 
If the ratio is unity,
${\mbox{\boldmath $\xi$}}$ turns out to be a homothetic vector. 
In the context of kinematic self-similarity, 
homothety is referred to as self-similarity of the first kind. 
If $\alpha=0$ and $\delta\ne 0$,
it is referred to as self-similarity of the zeroth kind. 
If the ratio is not equal to zero or one,
it is referred to as self-similarity of the second kind. 
If $\alpha\ne 0$ and $\delta=0$, 
it is referred to as self-similarity of the infinite kind.  
If $\delta=\alpha=0$, ${\mbox{\boldmath $\xi$}}$ turns out to be
a Killing vector.

From the Einstein equation (\ref{einstein}), we can derive
\begin{equation}
{\cal  L}_{\mbox{\boldmath $\xi$}} G_{\mu\nu} =8\pi G{\cal  L}_{\mbox{\boldmath $\xi$}} T_{\mu\nu}.
\label{intcondition}
\end{equation}
This equation is called the integrability condition. 
Now we can rewrite the integrability conditions (\ref{intcondition}) 
in terms of kinematic quantities of the fluid.
The covariant derivative of the fluid four velocity is decomposed into
the following form:
\begin{equation}
u_{\mu;\nu} =\sigma_{\mu\nu}+\frac{1}{3}\theta h_{\mu\nu}+\omega_{\mu\nu}
-\dot{u}_{\mu}u_{nu},
\end{equation}
where 
\begin{eqnarray}
\theta_{\mu\nu} &\equiv& 
h_{(\mu}^{~~\kappa}h_{\nu)}^{~~\lambda} u_{\kappa;\lambda}, \\
\theta &\equiv& g^{\mu\nu} \theta_{\mu\nu}, \\
\sigma_{\mu\nu} &\equiv& 
\theta_{\mu\nu}-\frac{1}{3}\theta h_{\mu\nu}, \\
\omega_{\mu\nu} &\equiv& 
h_{[\mu}^{~~\kappa}h_{\nu]}^{~~\lambda}u_{\kappa;\lambda}, \\
\omega^2 &\equiv& \frac{1}{2} \omega_{\mu\nu} \omega^{\mu\nu}, \\
\dot{u}_\mu &\equiv& u_{\mu;\nu} u^{\nu},
\end{eqnarray}
where the semicolon denotes the covariant derivative.
Using the above quantities, the integrability condition 
(\ref{intcondition}) is rewritten as follows~(cf.~\cite{coley1997}):
\begin{eqnarray}
&&(\delta-\alpha)(-8\omega^2 - 2 \dot{u}_\kappa^{~~;\kappa})= 8\pi G\left[\frac12({\cal L}_{\mbox{{\boldmath $\xi$}}}\mu + 2 \alpha \mu) + \frac32({\cal L}_{\mbox{{\boldmath $\xi$}}}p+2 \alpha p)\right],\label{int1}\\
&&2(\delta-\alpha)(\dot{\theta} + \theta^2 - 4 \omega^2)= 8\pi G\left[\frac32({\cal L}_{\mbox{{\boldmath $\xi$}}}\mu+2\delta\mu) -\frac32({\cal L}_{\mbox{{\boldmath $\xi$}}}p+2\delta p)\right],\label{int2}\\
&&2 \omega_{\lambda\mu} \dot{u}^\mu+2\omega_{\kappa\lambda}^{~~~;\kappa}-4\omega^2 u_\lambda=0,\label{int3} \\
&&\dot{\sigma}_{\lambda\rho}-u_\rho \sigma_{\lambda\nu} \dot{u}^\nu - u_\lambda \sigma_{\rho\mu} \dot{u}^\mu + \theta \sigma_{\lambda\rho} 
+ \sigma_{\lambda\kappa} \omega^\kappa_{~~\rho}+\sigma_{\rho\kappa} \omega^\kappa_{~~\lambda} + 2 \omega_\lambda^{~~\kappa} \omega_{\kappa\rho} + \frac{4}{3} h_{\lambda\rho}\omega^2 =0.\label{int4}
\end{eqnarray}
For the first-kind case, in which $\alpha=\delta\ne 0$, 
equations (\ref{ssp}) and (\ref{ssmu}) are obtained 
from equations (\ref{int1}) and (\ref{int2}). 
When a perfect fluid is irrotational, i.e., $\omega_{\mu\nu}=0$, 
the Einstein equations and the integrability conditions (\ref{int1})--(\ref{int4}) give~\cite{coley1997,mhio2003}
\begin{eqnarray}
(\alpha-\delta){\cal R}_{\mu\nu}=0,\label{3flat}
\end{eqnarray}
where ${\cal R}_{\mu\nu}$ is the Ricci tensor 
on the hypersurface orthogonal to $u^\mu$. 
This means that if a solution is kinematic self-similar but not
homothetic and if the fluid is irrotational,
then the hypersurface orthogonal to fluid flow is flat.

\section{Spherically symmetric self-similar solutions}
\subsection{Spherically symmetric solutions}
Although self-similar solutions can play important roles 
even in nonspherically symmetric solutions, such as 
homogeneous cosmological models~\cite{hw1990,we1997},
we focus in the rest of this article on spherically symmetric spacetimes. 
The line element in a spherically symmetric spacetime is given by 
\begin{eqnarray}
ds^2 = -e^{2\Phi(t,r)}dt^2+e^{2\Psi(t,r)}dr^2+R(t,r)^2 d\Omega^2,
\end{eqnarray}
where $d\Omega^2=d\theta^2+\sin^2 \theta d\varphi^2$. We consider a
perfect fluid as a matter field, for which the energy-momentum tensor is
given by equation (\ref{perfect_fluid}). 
We adopt the comoving coordinates, where the four-velocity of the fluid
$u^{\mu}$ has the components
\begin{equation}
u_{\mu} = (-e^{\Phi},0,0,0).
\end{equation}
Then, the Einstein equations and the equations of motion for the perfect fluid are reduced to the following simple form:
\bear
(\mu+p)\Phi_r &=& -p_r, \label{basic1}\\
(\mu+p)\Psi_t &=& -\mu_t-2(\mu+p)\frac{R_t}{R}, \label{basic2}\\
m_r &=& 4\pi \mu R_r R^2,   \label{basic3}\\
m_t &=& -4\pi p R_t R^2, \label{basic4}\\ 
0&=&-R_{tr}+\Phi_r R_{t}+\Psi_t R_r,\label{basic5}\\
2Gm &=& R(1+e^{-2\Phi}{R_t}^2-e^{-2\Psi}
{R_r}^2 ),\label{basic6}
\enar
where the subscripts $t$ and $r$ denote the partial derivatives 
with respect to $t$ and $r$, respectively, 
and $m(t,r)$ is called the Misner-Sharp mass. 
When a perfect fluid obeys an equation state $p+\mu=0$, which is
equivalent to a cosmological constant, the
first two equations are trivially satisfied.
In this case, one can use the following equations:
\begin{eqnarray}
&&-\frac{e^{2\Phi}}{R^2}-\left[\left(\frac{R_t}{R}\right)^{2}+
2\frac{R_t}{R} \Psi_t \right] +e^{2\Phi-2\Psi}\left[2\frac{R_{rr}}{R}-
2\frac{R_r}{R} \Psi_r+\left(\frac{R_r}{R}\right)^{2}\right]
=-8\pi G\mu e^{2\Phi},\label{00}\\
&&\frac{e^{2\Psi}}{R^2}+e^{2\Psi-2\Phi}\left[2\frac{R_{tt}}{R}-
2\frac{R_t}{R} \Phi_t+\left(\frac{R_t}{R}\right)^{2}\right] -\left[\left(\frac{R_r}{R}\right)^{2}+2\frac{R_r}{R} \Phi_r\right]
= -8\pi G p e^{2\Psi},\label{11}\\
&&e^{-2\Phi}\left({\Psi}_{tt}+{\Psi}_{t}^2
-\Phi_t \Psi_t+\frac{R_{tt}}{R}+ 
\frac{R_t \Psi_t}{R}-\frac{R_t \Phi_t}{R}\right) \nonumber \\
&&~~-e^{-2\Psi}\left(\Phi_{rr}+\Phi_r^2
-\Phi_r \Psi_r+\frac{R_{rr}}{R}
+\frac{R_r \Phi_r}{R}-\frac{R_r \Psi_r}{R}\right) 
=-8\pi G p, \label{22}
\end{eqnarray}
which are $(tt)$, $(rr)$ and $(\theta\theta)$ 
components of the Einstein equations, respectively. 
Five of the above nine equations are independent. 

\subsection{Spherically symmetric homothetic solutions}
There is a large variety of spherically symmetric homothetic 
solutions. The pioneering work in this area was done by Cahill and 
Taub~\cite{ct1971}.
The application contains
primordial black holes~\cite{ch1974,bh1978a,bh1978b},
cosmological 
voids~\cite{bertschinger1985,tomita1995,tomita1997a,tomita1997c}, 
cosmic 
censorship~\cite{newman1986,op1987,op1988a,wl1988,wl1989,op1990,lake1992,fh1993,hm2001} and 
critical behavior~\cite{choptuik1993,kha1995,ec1994}.
See~\cite{hin2002} and \cite{gundlach2003} for recent reviews of 
cosmic censorship and critical behavior, respectively.  
The classification of all spherically symmetric 
homothetic solutions with a perfect fluid has been 
made~\cite{gnu1998a,gnu1998b,carr2000,cc2000,ccgnu2000,ccgnu2001}.
The spacetime structure possible for homothetic solutions
has been studied~\cite{cg2002}.
The special case where the homothetic vector is orthogonal or parallel to the 
fluid flow has also been studied~\cite{mcintosh1975,coley1991}.
It has been revealed that a 
homothetic solution describes the dynamical properties of 
more general solutions in spherically symmetric gravitational  
collapse~\cite{hm2001}.
The stability of homothetic solutions has been 
studied ~\cite{kha1995,maison1996,kha1999,nc2000,hm2001,harada2001,bcgn2002,hm2004}.

When the spacetime admits a homothetic vector, which is 
neither parallel nor orthogonal to the fluid flow,
the homothetic vector ${\mbox{\boldmath $\xi$}}$ can be written as
\bear
{\mbox{\boldmath $\xi$}}=
t \frac{\partial}{\partial t}+r \frac{\partial}{\partial r},
\enar
and the self-similar variable $\xi$ is given by
\bear
\xi=\frac{r}{t}.
\enar
Homothety implies that the metric functions can be written
\bear
ds^2=-e^{2 \Phi(\xi)}dt^2+e^{2 \Psi(\xi)}dr^2+r^2 S(\xi)^2 d\Omega^2.
\enar
As we have seen, the equation of state must 
be of the form $p=K\mu$ for homothetic spacetimes. Then 
the governing equations for homothetic solutions are 
written as
\bear
e^{2 \Phi}&=&a_{\sigma}\xi^{\frac{4K}{1+K}}\eta^{-\frac{2K}{1+K}}, \\
e^{2 \Psi}&=&a_{\omega}\eta^{-\frac{2}{1+K}}S^{-4},\\
M+M^{\prime}&=&\eta S^{2}\left(S+S^{\prime}\right), \\
M^{\prime}&=&-K\eta S^{2}S^{\prime}, \\
\frac{M}{S}&=&1+a_{\sigma}^{-1}\left(\eta \xi^{-2}\right)^{\frac{2K}{1+K}} \xi^{2}S^{'2}-\eta^{\frac{2}{1+K}}S^{4}(S+S^{\prime})^{2},
\enar
where $a_{\sigma}$ and $a_{\omega}$ are integration constants and the prime denotes the derivative with respect to $\ln \xi$. The dimensionless functions $\eta(\xi)$ and $M(\xi)$ are defined by
\bear
8\pi G \mu&=&\frac{\eta}{r^2}, \\
2Gm&=&rM.
\enar
The above formulation is based on~\cite{ct1971,bh1978a,bh1978b}.
It is possible to choose another function 
in the same comoving coordinates,
as adopted in~\cite{ch1974,cc2000,cc2000b}. 
In the comoving coordinates, the dynamical properties of the 
fluid elements are very clear.

There are other useful formulations 
in analyzing homothetic solutions. 
One of the most natural coordinate systems for homothetic 
spacetimes is the so-called homothetic coordinates. 
In terms of this coordinate system, the dynamical systems theory
has been applied to homothetic solutions with a perfect fluid
for classification~\cite{bogoyavlenskii,gnu1998a,gnu1998b}.
In the homothetic coordinates, the self-similar variable
is chosen to be the spatial or time coordinate, depending
on whether the homothetic vector is timelike or spacelike.
If the homothetic vector is timelike, the line element 
is written as
\begin{eqnarray}
ds^2 = e^{2t}\left[-D_1^2(x)dt^2 + dx^2 + D_2^2(x)d\Omega^2\right].
\end{eqnarray}
If the homothetic vector is spacelike, the line element is written as
\begin{eqnarray}
ds^2 = e^{2x}\left[-dt^2 + D_1^2(t) dx^2 +
D_2^2(t)d\Omega^2\right].
\end{eqnarray}
If the homothetic vector is timelike in one region and
spacelike in another region of the same spacetime, 
the above two charts must be patched on the hypersurface 
on which the homothetic vector is null. 

Another coordinate system is that of area coordinates, in which 
the physical properties of the spacetime are clear.
The area coordinate system
has been adopted~\cite{op1987,op1988a,op1990}. 
In this coordinate system, the line element in homothetic spacetimes
is written as
\begin{eqnarray}
ds^2 &=&-e^{2{\bar \Phi}(z)}dt^2+e^{2{\bar \Psi}(z)}dr^2+r^2d\Omega^2,\\
z&=&\frac{r}{t}, \\
u^\mu \frac{\partial}{\partial x^\mu}&=&u^t(z)\frac{\partial}{\partial t}+u^r(z)\frac{\partial}{\partial r},
\end{eqnarray}
where $u^{t}$ and $u^{r}$ are also to be determined.

\subsection{Spherically symmetric kinematic self-similar solutions}
A kinematic self-similar vector may be parallel, orthogonal or tilted,
i.e., neither parallel nor orthogonal, to the fluid flow. 
Spherically symmetric kinematic self-similar perfect fluid solutions 
have been recently explored by several 
authors~\cite{bc1998,bc1999,sbc2001,
blvw2002,mhio2002a,mhio2002b,mhio2003}. 

In a spherically symmetric spacetime, the kinematic self-similar vector
field ${\mbox{\boldmath $\xi$}}$ is written in general as 
\begin{eqnarray}
{\mbox{\boldmath $\xi$}}=h_1(t,r) \frac{\partial}{\partial t}+h_2(t,r) \frac{\partial}{\partial r},
\end{eqnarray}
in the comoving coordinates,
where $h_1(t,r)$ and $h_2(t,r)$ are functions of $t$ and $r$. When 
$h_2=0$, ${\mbox{\boldmath $\xi$}}$ is parallel to the fluid flow, while when 
$h_1=0$, ${\mbox{\boldmath $\xi$}}$ is orthogonal to the fluid flow. 
When both $h_1$ and $h_2$ are nonzero,
${\mbox{\boldmath $\xi$}}$ is tilted. 

When the kinematic self-similar vector ${\mbox{\boldmath $\xi$}}$
is tilted to the fluid flow and not of the infinite kind,
${\mbox{\boldmath $\xi$}}$ and 
the metric tensor $g_{\mu\nu}$ 
are written in appropriate comoving coordinates as
\begin{eqnarray}
{\mbox{\boldmath $\xi$}}&=&(\alpha t+\beta) 
\frac{\partial}{\partial t}+r \frac{\partial}{\partial r}, \\
ds^2&=&-e^{2{\bar \Phi}(\xi)}dt^2+e^{2{\bar \Psi}(\xi)}dr^2+r^2 S(\xi)^2d\Omega^2,\label{finitessform}
\end{eqnarray}
where $\alpha$ is the index of self-similarity.
For $\alpha=1$, i.e., homothety or self-similarity of the first kind,
we can set $\beta=0$ and then 
$\xi$ is given by $\xi=r/t$.
For $\alpha=0$, i.e., self-similarity of the zeroth kind,
we can set $\beta=1$ and then 
$\xi$ is given by $\xi=r/e^{t}$.
For $\alpha\ne 0$ and $\alpha\ne1$, i.e. self-similarity of the second kind,
we can set $\beta=0$ and then 
$\xi$ is given by $\xi=r/(\alpha t)^{1/\alpha}$.
If the kinematic self-similar vector ${\mbox{\boldmath $\xi$}}$
is tilted to the fluid flow and of the infinite kind,
${\mbox{\boldmath $\xi$}}$ and 
the metric tensor $g_{\mu\nu}$ 
are written in appropriate comoving coordinates as
\begin{eqnarray}
{\mbox{\boldmath $\xi$}}&=&t\frac{\partial }{\partial t}
+r\frac{\partial }{\partial r}, \\
ds^2&=&-e^{2{\bar \Phi}(\xi)}dt^2+\frac{e^{2{\bar \Psi}(\xi)}}{r^2}dr^2+S(\xi)^2d\Omega^2\label{infinitessform},
\end{eqnarray}
where the self-similar variable is given by $\xi=r/t$.

If the kinematic self-similar vector ${\mbox{\boldmath $\xi$}}$ 
is parallel to the fluid flow and not of the infinite kind, 
we have in appropriate comoving coordinates as
\begin{eqnarray}
{\mbox{\boldmath $\xi$}}&=&t \frac{\partial}{\partial t},\\
ds^2 &=& -t^{2(\alpha-1)}e^{2{\bar \Phi}(r)}dt^2
+t^2dr^2+t^2 S(r)^2 d\Omega^2, \label{parafinite}
\end{eqnarray}
where $\alpha$ is the index of self-similarity and 
the self-similar variable is given by $\xi=r$. 
If the kinematic self-similar vector ${\mbox{\boldmath $\xi$}}$ 
is parallel to the fluid flow and of the infinite kind, 
we have
in appropriate comoving coordinates as
\begin{eqnarray}
{\mbox{\boldmath $\xi$}}&=&t \frac{\partial}{\partial t},\\
ds^2 &=& -e^{2{\bar \Phi}(r)}dt^2+dr^2+S(r)^2 d\Omega^2, \label{parainfinite}
\end{eqnarray}
where the self-similar variable is given by $\xi=r$. 

If the kinematic self-similar vector ${\mbox{\boldmath $\xi$}}$ is
orthogonal to the fluid flow and not of the infinite kind, we have
in appropriate coordinates
\begin{eqnarray}
{\mbox{\boldmath $\xi$}} &=&r \frac{\partial}{\partial r},\\
ds^2 &=& -r^{2\alpha}dt^2+e^{2{\bar \Psi}(t)}dr^2+r^2 
S(t)^2 d\Omega^2 \label{orthfinite},
\end{eqnarray}
where $\alpha$ is the index of self-similarity and 
the self-similar variable is given by $\xi=t$.
If the kinematic self-similar vector ${\mbox{\boldmath $\xi$}}$ is
orthogonal to the fluid flow and of the infinite kind, we have
in appropriate coordinates
\begin{eqnarray}
{\mbox{\boldmath $\xi$}} &=&r \frac{\partial}{\partial r},\\
ds^2 &=& -r^2dt^2+\frac{e^{2{\bar \Psi}(t)}}{r^2}dr^2+S(t)^2 d\Omega^2,
\end{eqnarray}
where the self-similar variable is given by $\xi=t$.

Not as homothetic solutions in the tilted case, 
kinematic self-similar solutions have a characteristic structure. 
We now show an example of them in the case of self-similarity of 
the second kind, where a kinematic self-similar vector is 
tilted to the fluid flow. 
In this case, the Einstein equations imply that 
the quantities $m$, $\mu$ and $p$ must be of the following form: 
\begin{eqnarray}
\frac{2Gm}{r}&=& M_1(\xi)+\frac{r^2}{t^2}M_2(\xi),\label{secondm}\\
8\pi G \mu r^{2}&=&W_1(\xi)+\frac{r^2}{t^2}W_2(\xi),\label{secondmu}\\
8\pi G pr^{2}&=&P_1(\xi)+\frac{r^2}{t^2}P_2(\xi),\label{secondp}
\end{eqnarray}
where $\xi=r/(\alpha t)^{1/\alpha}$. 
In other words, dimensionless quantities on the left hand side 
are decomposed into two parts, one remains constant and 
the other behaves as $(r/t)^{2}\propto r^{2(1-\alpha)}$
as $\xi$ is fixed. 
Then, the original partial differential equations are satisfied
when and only when the Einstein equations and 
the equations of motion for the matter field are satisfied 
for each of the $O(1)$ and $O[(r/t)^2]$ terms. 
The equations (\ref{basic1})--(\ref{11})  
for a perfect fluid 
then reduce to the following:
\begin{eqnarray}
M_1+M_1'&=&W_1S^2(S+S'),\label{second1}\\
3M_2+M_2'&=&W_2S^2(S+S'),\label{second2}\\
M_1'&=&-P_1S^2S',\label{second3}\\
2\alpha M_2+M_2'&=&-P_2S^2S',\label{second4}\\
M_1&=&S[1-e^{-2\Psi}(S+S')^2],\label{second5}\\
\alpha^2 M_2&=&SS^{'2} e^{-2\Phi},\label{second6}\\
(P_1+W_1)\Phi'&=&2P_1-P_1',\label{second7}\\
(P_2+W_2)\Phi'&=&-P_2',\label{second8}\\
W_1'S&=&-(P_1+W_1)(\Psi'S+2S'),\label{second9}\\
(2\alpha W_2+W_2')S&=&-(P_2+W_2)(\Psi'S+2S'),\label{second10}\\
S''+S'&=&S'\Phi'+(S+S')\Psi',\label{second11}\\
S'(S'+2\Psi'S)&=&\alpha^2W_2 S^2 e^{2\Phi}, \label{second00a}\\
2S(S''+2S')-2\Psi'S(S+S')&=&-S^{'2}-S^2+e^{2\Psi}(1-W_1S^2),\label{second00b}\\
2S(S''+\alpha S'-\Phi'S')+S^{'2}&=&-\alpha^2 P_2S^2 e^{2\Phi},\label{second11a}\\
(S+S')(S+S'+2\Phi'S)&=&(1+P_1S^2)e^{2\Psi},\label{second11b}
\end{eqnarray}
where we have omitted the bars of ${\bar \Phi}$ and ${\bar \Psi}$ in (\ref{finitessform}) for simplicity and the prime denotes the derivative with respect to $\ln \xi$. A
similar structure of basic equations can be found for kinematic
self-similar solutions of the second, zeroth and infinite kinds both in
the tilted and orthogonal cases and of the second and zeroth kind in the
parallel case. The exceptions are the first kind in the tilted, parallel and orthogonal cases and the infinite kind in the parallel
case. 
See~\cite{mhio2002b,mhio2003} 
for the basic equations for spherically symmetric self-similar 
solutions for all cases.

It is interesting to consider the 
spherically symmetric self-similar solutions of the infinite kind
with a kinematic self-similar vector parallel to the 
fluid flow. 
The metric form demanded by this self-similarity, which is 
given by equation (\ref{parainfinite}),
is nothing but the general form of the line element 
in spherically symmetric static spacetimes when the chosen radial 
coordinate is the radial physical length.
Therefore, all static solutions have a kinematic self-similar
vector of the infinite kind that is parallel to the fluid flow. 
Inversely, all spherically symmetric solutions 
with a kinematic self-similar vector of the infinite kind 
parallel to the fluid flow are static.
The equation of state is not restricted at all.

\subsection{Equation of state}
It is obvious that equations (\ref{secondmu}) 
and (\ref{secondp}) strongly restrict the
form of the possible equations of state.
The detailed analysis shows the following restriction on the 
equation of state~\cite{mhio2002b}.
Suppose we have the barotropic equation of state, i.e., $p=f(\mu)$.
The self-similarity of the second kind with the index $\alpha$
can admit only the following 
equation of state:
\begin{equation}
k_{1}x+k_{2}x^{\alpha}=f(C_{1}x+C_{2}x^{\alpha}),
\end{equation}
where $k_{1}$, $k_{2}$, $C_{1}$ and $C_{2}$ are 
arbitrary constants.
For the self-similarity of the zeroth and infinite kinds,
we cannot determine the equation of state alone from the 
decomposed form.
It should be noted that the above discussion is based
not on the whole equations but only on the decomposed form 
of $p$ and $\mu$ such as equations (\ref{secondmu}) 
and (\ref{secondp}).

For later convenience, we introduce 
and focus on the following equations of state:
\begin{itemize}
\item 
Equation of state (1) (EOS1)
\bear
p=K \mu^{\gamma}, \label{eos1}
\enar
where $K$ and $\gamma$ are constants. Here we assume that $K \ne 0$ and 
$\gamma \ne 0,1$,
\item 
Equation of state (2) (EOS2)
\bear
\left\{
\begin{array}{ll}
\displaystyle{p=K n^{\gamma}},\\
\displaystyle{\mu=m_b n+\frac{p}{\gamma-1}},\label{eos2}
\end{array}
\right.
\enar
where the constant $m_b$ and $n(t,r)$ correspond to the mean baryon mass and the baryon number density, respectively. Here we assume that $K \ne 0$ and $\gamma \ne 0,1$.
In the literature, this equation of state is sometimes called a
relativistic polytrope,
\item 
Equation of state (3) (EOS3)
\bear
p=K \mu. \label{eos3}
\enar
where we assume that $-1\le K \le 1$.
\end{itemize}
EOS1 and EOS2 are two kinds of polytropic equations of state. These
equations of state are incompatible with homothety. 
For $0<\gamma<1$, both EOS1 and EOS2 are approximated by
a dust fluid in the high-density regime.
For $1<\gamma$, EOS2 is approximated by EOS3 with $K=\gamma-1$
in the high-density regime.
For $2<\gamma$ for EOS2 and $1<\gamma$ for EOS1, 
the dominant energy condition can be violated in the 
high-density regime, which would be unphysical.

\setcounter{equation}{0}
\section{Exact spherically symmetric self-similar 
solutions}\label{sec3}
\setcounter{equation}{0}
\subsection{Vacuum}
In a vacuum, the only spherically symmetric solutions are the Minkowski solution and
the Schwarzschild solution from Birkhoff's theorem. Both 
solutions have kinematic self-similar vectors. Although there are no
fluids, we can introduce a unit timelike vector
$u^\mu$.

The Minkowski solution has seven kinematic self-similar vectors including
a homothetic  vector in the tilted case.
The Minkowski solution is represented by
\bear
ds^2&=&-dt^2+dr^2+r^2d\Omega^2,\label{minkowski}\\
2Gm&=&0,\\
8\pi G p&=&8\pi G \mu=0.
\enar
The metric can be represented in the Milne form
\bear
ds^2=-d\tau^2+\tau^2d\rho^2+\tau^{2}\sinh^2 \rho d\Omega^2,
\enar
where $t=\tau \cosh \rho$ and $r= \tau \sinh \rho$, or in another 
form
\bear
ds^2=-\varpi^2d\nu^2+d\varpi^2+\varpi^2\cosh^2 \nu d\Omega^2,
\enar
where $t=\varpi \sinh \nu$ and $r= \varpi \cosh \nu$.
This spacetime has the following kinematic self-similar vectors: 
\begin{itemize}
\item First kind, tilted\\
\bear
t\frac{\partial}{\partial t}+r\frac{\partial}{\partial r},
\enar
\item First kind, parallel\\
\bear
\tau\frac{\partial}{\partial \tau},
\enar
\item First kind, orthogonal\\
\bear
\varpi\frac{\partial}{\partial \varpi},
\enar
\item Second kind with any $\alpha$, tilted\\
\bear
\alpha t\frac{\partial}{\partial t}+r\frac{\partial}{\partial r},
\enar
\item Zeroth kind, tilted\\
\bear
\frac{\partial}{\partial t}+r\frac{\partial}{\partial r},
\enar
\item Zeroth kind, orthogonal\\
\bear
r\frac{\partial}{\partial r},
\enar
\item Infinite kind, parallel\\
\bear
t\frac{\partial}{\partial t}.
\enar
\end{itemize}

The Schwarzschild
solution has two kinematic self-similar vectors but does not have a
homothetic vector.
The Schwarzschild solution is represented by
\bear
ds^2&=&-\left(1-\frac{2Gm_0}{r}\right)dt^2+\frac{dr^2}{1-\displaystyle{
\frac{2Gm_0}{r}}}+r^2d\Omega^2,\\
2Gm&=&2Gm_0,\\
8\pi G p&=&8\pi G \mu=0,
\enar
where $m_0$ is a constant. This spacetime can be
represented in the following choice of coordinates:
\begin{equation}
ds^2=-d\tau^2+(2Gm_0)^{2/3}\left(\frac{d\rho^2}
{\left[\frac32(\rho-\tau)\right]^{2/3}}+
\left[\frac32(\rho-\tau)\right]^{4/3}d\Omega^2\right).
\end{equation}
This spacetime has the following kinematic self-similar vectors: 
\begin{itemize}
\item Second kind with $\alpha=3/2$, tilted\\
\bear
\tau\frac{\partial}{\partial \tau}+\rho \frac{\partial}{\partial \rho},
\enar
\item Infinite kind, parallel\\
\bear
t\frac{\partial}{\partial t}.
\enar
\end{itemize}

\subsection{Cosmological constant}
Since the cosmological constant
introduces a length scale $1/\sqrt{|\Lambda|}$,
solutions cannot be homothetic. 
However, the de Sitter solution, the Schwarzschild-de Sitter solution and the Nariai solution 
admit kinematic self-similar vectors.
 
The de Sitter solution is represented by
\bear
ds^2&=&-dt^2+e^{2\sqrt{\Lambda/3}t}(dr^2+r^2d\Omega^2),\\
2Gm&=&\frac{\Lambda}{3}r^3e^{3\sqrt{\Lambda/3}t},\\
8\pi G p&=&-8\pi G \mu=-8\pi G\Lambda,
\enar
where $\Lambda$ is a cosmological constant. 
This solution is represented in the static coordinates as
\bear
ds^2&=&-\left(1-\frac13\Lambda \rho^2\right)d\tau^2+
\left(1-\frac13\Lambda \rho^2\right)^{-1}d\rho^2+\rho^2d\Omega^2,\\
2Gm&=&\frac13\Lambda \rho^3,\\
8\pi G p&=&-8\pi G \mu=-8\pi G\Lambda.
\enar
When $\Lambda$ is negative, the solution is called the 
anti de Sitter solution. 
The de Sitter solution has the following kinematic self-similar vectors: 
\begin{itemize}
\item Zeroth kind, tilted\\
\bear
\frac{\partial}{\partial t}+\lambda r \frac{\partial}{\partial r},
\enar
where $\lambda$ is a non-zero constant.
\item Zeroth kind, parallel\\
\bear
\frac{\partial}{\partial t},
\enar
\item Zeroth kind, orthogonal\\
\bear
r \frac{\partial}{\partial r},
\enar
\item Infinite kind, parallel\\
\bear
\tau \frac{\partial}{\partial \tau}.
\enar
\end{itemize}

The Schwarzschild-de Sitter solution is an exact solution with a cosmological 
constant, which is represented by
\bear
ds^2&=&-\left(1-\frac{2Gm_0}{\rho}-\frac13\Lambda \rho^2\right)d\tau^2
+\left(1-\frac{2Gm_0}{\rho}-\frac13\Lambda \rho^2\right)^{-1}d\rho^2
+\rho^2d\Omega^2,\\
2Gm&=&2Gm_0+\frac13\Lambda \rho^3,\\
8\pi G p&=&-8\pi G \mu=-8\pi G\Lambda.
\enar
where $m_0$ is a constant and $\Lambda$ is a cosmological constant. 
When $\Lambda$ is negative, the solution is called the 
Schwarzschild-anti de Sitter solution. 
The Schwarzschild-de Sitter solution has the 
following kinematic self-similar vector: 
\begin{itemize}
\item Infinite kind, parallel\\
\bear
\tau \frac{\partial}{\partial \tau}.
\enar
\end{itemize}

The Nariai solution~\cite{nariai} is an exact solution with a cosmological constant,
which is represented by
\bear
ds^2&=&-\left[a(t)\sin\left(\ln(\sqrt{\Lambda}r)\right)+b(t)\cos\left(\ln(\sqrt{\Lambda}r)\right)\right]^2dt^2+\frac{1}{\Lambda r^2}(dr^2+r^2d\Omega^2),\\
2Gm&=&1/\sqrt{\Lambda},\\
8\pi G p&=&-8\pi G\mu=-\Lambda,
\enar
where $a$ and $b$ are arbitrary functions of $t$. With the choice 
\begin{eqnarray}
a&=&\frac{t^{1/(c_1\sqrt{\Lambda})-1)}}{c_1\sqrt{\Lambda}}
(A\cos(\ln t)+B\sin(\ln t)),\\
b&=&\frac{t^{1/(c_1\sqrt{\Lambda})-1}}{c_1\sqrt{\Lambda}}
(-A\sin(\ln t)+B\cos(\ln t)), 
\end{eqnarray}
where $A$ and $B$ are constants, and the coordinate transformation 
\begin{equation}
r=\frac{r'}{\sqrt{\Lambda}}, 
\end{equation}
this metric is written as
\bear
ds^2=-\frac{1}{c_1^2\Lambda}t^{2/(c_1\sqrt{\Lambda})-2}\left[A \sin\left(\ln \frac{r'}{t}\right)+B \cos\left(\ln\frac{r'}{t}\right)\right]^2dt^2+\frac{1}{\Lambda r^{'2}}(dr^{'2}+r^{'2}d\Omega^2). \label{nariailike1}
\enar
This spacetime is also written in the static coordinates as
\bear
ds^2=-\left[{\bar A} \sin\left(\ln (\sqrt{\Lambda}r)\right)+{\bar B} \cos\left(\ln (\sqrt{\Lambda}r)\right)\right]^2d\tau^2+\frac{1}{\Lambda r^2}(dr^2+r^2d\Omega^2),
\enar
where ${\bar A}$ and ${\bar B}$ are constants. 
This spacetime has the following kinematic self-similar vectors: 
\begin{itemize}
\item Infinite kind, tilted\\
\bear
t\frac{\partial}{\partial t}+r'\frac{\partial}{\partial r'},
\enar
\item Infinite kind, parallel\\
\bear
\tau\frac{\partial}{\partial \tau}.
\enar
\end{itemize}

\subsection{Dust fluid}
Without assumption of self-similarity, the general solution for a
spherically symmetric dust fluid is exactly obtained, which is called 
the Lema\^{\i}tre-Tolman-Bondi (LTB) 
solution~\cite{lemaitre1933,tolman1934,bondi1947}.

Therefore, spherically symmetric self-similar solutions with a dust 
fluid are a subclass of the LTB solutions.
Homothetic solutions in the tilted case was
completely classified~\cite{carr2000}. 
The homothetic LTB solutions have been discussed in the 
context of cosmic censorship~\cite{newman1986}.
The homothetic LTB solution is represented by
\bear
ds^2&=&-dt^2+\frac{(S+\xi S_\xi)^2}{1+2E}dr^2+r^2S^2d\Omega^2,\\
2Gm&=&2\sqrt{1+2E} r,\\
8\pi G \mu&=&\frac{2\xi\Gamma}{r^2S^2(\xi S\pm\sqrt{2E+2\Gamma/S})},
\enar
where $\xi=r/t$, 
where a subscript $\xi$ means the derivative with respect to $\xi$, and where $E$ and $\Gamma$ are constants with a relation
\bear
E=\frac12(\Gamma^2-1).
\enar
$S$ is given by 
\begin{equation}
D \mp \frac{1}{\xi}= \left\{
\begin{array}{ll}
\displaystyle{\frac{\sqrt{ES^2+\Gamma S}}{\sqrt{2} E}
-\frac{2\Gamma}{(2E)^{\frac{3}
{2}}}\sinh^{-1}{\sqrt{\frac{ES}{\Gamma}}}} &\mbox{for}
\quad E>0,\\
\displaystyle{\frac{\sqrt{2}}{3}S^{\frac{3}{2}}} &
\mbox{for}\quad E=0,\\
\displaystyle{\frac{2\Gamma}{(-2E)^{\frac32}}\sin^{-1}
\sqrt{\frac{-ES}{\Gamma}} \pm 
\frac{\sqrt{ES^2+\Gamma S}}{\sqrt{2}E}} &
\mbox{for}\quad -\frac12<E<0,
\end{array} \right. 
\end{equation}
where $D$ is a constant. 
This solution has a homothetic  vector
\bear
t\frac{\partial}{\partial t}+r\frac{\partial}{\partial r}.
\enar
$E$ can be interpreted as the sum of the kinetic energy and 
the potential energy per unit mass. When $E=0$, each shell is 
marginally bound.
This solution is a two-parameter family of solutions of $E$ and $D$ and
reduces to the flat Friedmann-Robertson-Walker (FRW) solution when
$E=D=0$. 
It is noted that there are nonmarginally bound LTB solutions 
with a homothetic vector. 

In addition, there are kinematic self-similar solutions
of the second, zeroth and infinite kinds. From
equation~(\ref{3flat}), any three-surface orthogonal to the fluid flow
in a kinematic self-similar solution to the Einstein equations of the
second, zeroth or infinite kind that contains only irrotational dust
as a matter field is flat. Because the
flatness of the three-surface implies that these solutions are 
marginally bound, a spherically symmetric kinematic
self-similar solution to the Einstein equations of the second, zeroth or
infinite kind that contains only a dust fluid 
is described by the marginally bound LTB solutions. 
These solutions have been investigated by several 
authors~\cite{ch1989,bc1998,blvw2002,mhio2003}.

The second-kind kinematic self-similar 
LTB solution is represented by
\bear
ds^2&=&-d t^2+\frac{9[\kappa r^{3\alpha/(3-2\alpha)}+(2\alpha/3-1) t]^2}{(3-2\alpha)^2|\kappa r^{3\alpha/(3-2\alpha)}- t|^{2/3}}d r^2+ r^2|\kappa r^{3\alpha/(3-2\alpha)}- t|^{4/3}d\Omega^2,\label{solsecond1}\\
8\pi G\mu&=&\frac{4(3-2\alpha)}{9[\kappa r^{3\alpha/(3-2\alpha)}+(2\alpha/3-1) t](\kappa r^{3\alpha/(3-2\alpha)}- t)},\label{solsecond2}\\
2Gm&=&\frac49  r^3, \label{solsecond3}
\enar
where $\kappa$ is an arbitrary dimensional constant and $\alpha\ne
3/2$.
For $\alpha=3/2$, the solution turns out to be the flat FRW solution.
This spacetime has the following kinematic self-similar vector: 
\bear
\alpha t \frac{\partial}{\partial t}+\frac{3-2\alpha}{3}r \frac{\partial}{\partial r}.
\enar

The zeroth-kind kinematic self-similar LTB solution is represented by
\bear
ds^2&=&-d t^2+\frac{( t-2 \lambda /3-\lambda \ln  r)^2}{| t-\lambda 
\ln  r|^{2/3}}d r^2+ r^2| t-\lambda \ln  r|^{4/3}d\Omega^2,\label{solzero1}\\
8\pi G\mu&=&\frac{4}{3( t-\lambda \ln  r)( t-2\lambda /3-\lambda \ln  r)},\label{solzero2}\\
2Gm&=&\frac49  r^3. \label{solzero3}
\enar
This spacetime has the following kinematic self-similar vector: 
\bear
\lambda \frac{\partial}{\partial t}+r \frac{\partial}{\partial r},
\enar
where $\lambda $ is an arbitrary dimensional constant.

The infinite kind kinematic self-similar LTB solution is represented by
\bear
ds^2&=&-d t^2+ t^{2}|\sigma r^{-3/2}- t|^{-2/3}d r^2+ r^2|\sigma r^{-3/2}- t|^{4/3}d\Omega^2,\label{infinite1}\\
8\pi G\mu&=&\frac{-4}{3 t(\sigma r^{-3/2}- t)},\label{infinite2}\\
2Gm&=&\frac49  r^3. \label{infinite3}
\enar
This spacetime has the following kinematic self-similar vector: 
\bear
t \frac{\partial}{\partial t}-\frac23 r \frac{\partial}{\partial r},
\enar
where $\sigma$ is an arbitrary dimensional constant.

The flat FRW 
solution with a dust fluid will be discussed 
together with those with a perfect fluid.

\subsection{Perfect fluid}
\subsubsection{Homothetic solutions} 
As examples of homothetic exact solutions with a perfect fluid
obeying the equation of state $p=K\mu$, we discuss 
the power-law flat FRW solution, 
the homothetic static perfect fluid solution,
and the Kantowski-Sachs solution. 
The power-law flat FRW solution and the homothetic static solution
have kinematic self-similar vectors as well as a homothetic vector.
The Kantowski-Sachs solution has no kinematic self-similar vector 
except a homothetic vector.

The flat FRW solution has the following form:
\bear
ds^2&=&-dt^2+t^{[4/(3(1+K)]}(dr^2+r^2d\Omega^2),\\
2Gm&=&\frac{4}{9(1+K)^2}r^3t^{-2K/(1+K)},\\
8\pi G p&=&8\pi G K\mu=\frac{4K}{3(1+K)^2t^2}.
\enar
This solution has the following kinematic self-similar vectors:
\begin{itemize}
\item First kind, tilted ($K\ne -1/3$)\\
\bear
t\frac{\partial}{\partial t}+\frac{1+3K}{3(1+K)}r\frac{\partial}{\partial r},
\enar
\item First kind, parallel ($K=-1/3$)\\
\bear
t \frac{\partial}{\partial t},
\enar
\item Second kind for $\alpha \ne 3(1+K)/2$, tilted\\
\bear
\alpha t \frac{\partial}{\partial t}+\left(1-\frac{2\alpha}{3(1+K)}\right) r \frac{\partial}{\partial r},
\enar
\item Second kind with $\alpha=3(1+K)/2$, parallel\\
\bear
\alpha t \frac{\partial}{\partial t},
\enar
\item Zeroth kind, orthogonal\\
\bear
r \frac{\partial}{\partial r},
\enar
\item Infinite kind, tilted\\
\bear
t \frac{\partial}{\partial t}
-\frac{2}{3(1+K)}r \frac{\partial}{\partial r}.
\enar
\end{itemize}

The homothetic static perfect fluid solution is represented as the following: 
\bear
ds^2&=&-r^{4K/(1+K)} dt^2+\frac{K^2+6K+1}{(1+K)^2}dr^2+r^2 d\Omega^2,\\
2Gm&=&\frac{4K}{K^2+6K+1}r,\\
8\pi G p&=&8\pi G K\mu=\frac{4K^2}{(K^2+6K+1)r^2}.
\enar
Since the center $r=0$ is singular and timelike, it must be 
a naked singularity.
This solution has the following kinematic self-similar vectors:
\begin{itemize}
\item First kind, tilted ($K\ne 1$)\\
\bear
\frac{1-K}{1+K}t\frac{\partial}{\partial t}
+r\frac{\partial}{\partial r}.
\enar
\item First kind, orthogonal ($K= 1$)\\
\bear
r\frac{\partial}{\partial r},
\enar
\item Second kind for $\alpha\ne 2K/(1+K)$, tilted\\
\bear
\left(\alpha-\frac{2K}{1+K}\right)
t\frac{\partial}{\partial t}+r\frac{\partial}{\partial r},
\enar
\item Second kind with $\alpha=2K/(1+K)$, orthogonal\\
\bear
r\frac{\partial}{\partial r},
\enar
\item Zeroth kind, tilted\\
\bear
-\frac{2K}{1+K}t\frac{\partial}{\partial t}
+r\frac{\partial}{\partial r}.
\enar
\end{itemize}

The Kantowski-Sachs solution is represented by
\bear
ds^2&=&-dt^2+t^{-4K/(1+K)}dr^2+\frac{(1+K)^2}{(1+3K)(K-1)}t^2d\Omega^2,\\
2Gm&=&\frac{4(1+K)K^2t}{(1+3K)^{3/2}(K-1)^{3/2}},\\
8\pi G p&=&8\pi G K\mu=-\frac{4K^2}{(1+K)^2t^2},
\enar
with a homothetic vector
\bear
t\frac{\partial}{\partial t}+\frac{1+3K}{1+K}r\frac{\partial}{\partial r}.
\enar
$-1<K<-1/3$ must be satisfied for this solution to be physical
and in that case the above homothetic vector is tilted.

\subsubsection{Nonhomothetic kinematic self-similar solutions}
As examples of nonhomothetic kinematic self-similar 
solutions, we discuss 
the general FRW solutions and 
the Gutman-Bespal'ko solution.

The general FRW solution has kinematic self-similar vectors and 
does not have homothetic vectors if it is not power-law or 
if it is not flat. 
The general FRW solutions are given by
\bear
ds^2&=&-dt^2+a(t)^2(dr^2+S(r)^2d\Omega^2),\\
2Gm&=&aS(1-S^{'2}+{\dot a}^2S^2),\\
8\pi G p&=&-2\frac{{\ddot a}}{a}-\left(\frac{{\dot a}}{a}\right)^2-\frac{k}{a^2},\\
8\pi G \mu&=&3\left(\frac{{\dot a}}{a}\right)^2+\frac{3k}{a^2},\\
S(r)&=&\left\{
\begin{array}{ll}
\sin{r}, &\quad \mbox{for}\quad k=1 \\
r, &\quad \mbox{for}\quad k=0 \\
\sinh{r},&\quad \mbox{for} \quad k=-1 \\ 
\end{array}\right.
\enar
where a dot and a prime denote the derivatives with respect to $t$ and $r$, respectively. 

The non-power-law flat FRW solution ($k=0$) has 
the following kinematic self-similar vector 
independent of the form of the equation of state:
\begin{itemize}
\item Zeroth kind, orthogonal\\
\bear
r \frac{\partial}{\partial r}.
\enar
\end{itemize}

The closed FRW solution ($k=1$) has the
following kinematic self-similar vector for EOS2
with $\gamma=2/3$:
\begin{itemize}
\item Second kind with $\alpha=3/2$, parallel\\
\bear
t\frac{\partial}{\partial t}.
\enar
\end{itemize}

The curved FRW solutions ($k=\pm 1$) 
with the equation of state $p=-\mu/3$ 
have the following additional kinematic self-similar vector:
\begin{itemize}
\item First kind, parallel\\
\bear
t \frac{\partial}{\partial t}+r \frac{\partial}{\partial r}.
\enar
\end{itemize}

Only for a stiff fluid $p=\mu$, the Gutman-Bespal'ko solution
exists, which
is represented by~\cite{gb1967,exactsolutions2003,mhio2003} 
\bear
ds^2&=&-\frac{1}{4}r^2dt^2+dr^2+\frac{1}{2}r^2
(1+a_{1}e^{t}+a_{2}e^{-t})
d\Omega^2,\\
2Gm&=&\frac{1}{2\sqrt{2}}(1-4a_{1}a_{2})(1+a_{1}e^{t}+a_{2}e^{-t})^{-3/2}r,\\
8\pi G p&=&8\pi G \mu=(1-4a_{1}a_{2})(1+a_{1}e^{t}+a_{2}e^{-t})^{-2}r^{-2},
\enar
where $a_{1}$ and $a_{2}$ are arbitrary constants.
In this spacetime the physical center $r=0$ is singular. 
This spacetime 
has the following kinematic self-similar vector:
\begin{itemize}
\item First kind, orthogonal\\
\bear
r\frac{\partial}{\partial r}.
\enar
\end{itemize}
This solution includes the homothetic static 
perfect fluid solution for a stiff fluid as a special case 
$a_{1}=a_{2}=0$.

The results in this section are summarized in Table~\ref{tablesol4}.
As we will see in section~\ref{sec:nonexistence}, it can be shown that 
this table provides the complete list of kinematic self-similar
solutions compatible with EOS1, EOS2 and EOS3.

\section{Nonexistence of kinematic self-similar solutions 
with a polytropic equation of state}
\label{sec:nonexistence}
Among kinematic self-similarities, homothety in the tilted
case includes a large variety of solutions and has been intensively
investigated. As we have seen, the equation of state is restricted
to be of the form $p=K\mu$ for homothetic solutions.

In this section, based on~\cite{mhio2002a,mhio2002b,mhio2003},
we will briefly see that there are no kinematic self-similar
solutions with nontrivial polytropic equations of state although
we first expected that the generalization from homothety to kinematic
self-similarity would enable us to analyze a wider class of 
physical solutions.
As we have already mentioned, it is possible to construct a
characteristic length scale from given dimensional constants in the
general relativistic system of a perfect fluid with a polytropic
equation of state. According to a usual procedure, we have introduced
incomplete similarity into this system. Kinematic self-similarity is a
natural generalization of incomplete similarity into general relativity.
Therefore, it is highly nontrivial that there are no 
kinematic self-similar solutions with a kinematic self-similar 
vector tilted to the flow of a perfect 
fluid with a polytropic equation of state.

\subsection{Tilted cases}
\subsubsection{Second kind}
We consider the second-kind kinematic self-similar solutions in the 
tilted case, in which the energy density and the pressure of a perfect fluid 
are of the form (\ref{secondmu}) and (\ref{secondp}), respectively. The 
equation of state gives the relation among the functions $P_1$, $P_2$, 
$W_1$ and $W_2$. EOS1 admits the following two cases:
\begin{eqnarray}
&&\alpha=\gamma, \quad P_1=W_2=0,\quad P_2 =\frac{K}{(8\pi G)^{\gamma-1}
\gamma^2}{\xi}^{-2\gamma}W_1^{\gamma}, \quad \mbox{(A)} \
\label{casea}\\
&&\alpha=\frac{1}{\gamma},\quad P_2=W_1=0,\quad P_1 =\frac{K}{(8\pi G)^
{\gamma-1}\gamma^{2\gamma}}{\xi}^2 W_2^{\gamma}, \quad \mbox{(B)} 
\label{caseb}
\end{eqnarray}
while EOS2 admits the following two cases:
\begin{eqnarray}
&&\alpha=\gamma,\quad P_1=0,\quad P_2 =\frac{K}{m_b^{\gamma}(8\pi G)^{\gamma-1}\gamma^2}{\xi}^{-2\gamma}W_1^{\gamma}=(\gamma-1)W_2, \quad \
\mbox{(C)} \label{casec}\\
&&\alpha=\frac{1}{\gamma},\quad P_2=0,\quad P_1 =\frac{K}{m_b^{\gamma}
(8\pi G)^{\gamma-1}\gamma^{2\gamma}}{\xi}^2 W_2^{\gamma}=(\gamma-1)W_1. 
\quad \mbox{(D)} \label{cased}
\end{eqnarray}
We can show that none of these cases satisfies the 
Einstein equations although they are compatible with the 
decomposition given by equations (\ref{secondmu}) and (\ref{secondp}).
Subtracting equation (\ref{second11b}) from (\ref{second00b}) and 
eliminating 
$S''$ by use of equation (\ref{second11}), we obtain
\begin{eqnarray}
2\Phi'=(P_1+W_1)e^{2\Psi}. \label{2ndkey1}
\end{eqnarray}
Then equations (\ref{second7}) and (\ref{second8}) result in
\begin{eqnarray}
e^{2\Psi}(P_1+W_1)^2&=&4P_1-2P_1',\label{2ndkey2}\\
e^{2\Psi}(P_1+W_1)(P_2+W_2)&=&-2P_2'.\label{2ndkey3}
\end{eqnarray}
It is obvious that $P_1=0$ implies $W_1=0$ from equation (\ref{2ndkey2}) 
while
$P_2=0$ implies $W_2=0$ or $P_{1}+W_{1}=0$
from equation (\ref{2ndkey3}). Therefore, it is 
concluded 
that all cases (A)--(D) result in vacuum spacetimes.  

\subsubsection{Zeroth kind}
Next, we consider the zeroth-kind kinematic self-similar solutions in 
the tilted case. In this case, the Einstein equations imply that the 
quantities $\mu, p$ and $m$ are of the forms 
\begin{eqnarray}
\frac{2Gm}{r}&=&M_1(\xi)+r^2 M_2(\xi),\label{zerom}\\
8\pi G \mu r^2&=&W_1(\xi)+r^2W_2(\xi),\label{zeromu}\\
8\pi G p r^2&=&P_1(\xi)+r^2P_2(\xi),\label{zerop}
\end{eqnarray}
where $\xi=r/e^{t}$. A set of ordinary differential equations is 
obtained when it is stipulated that the Einstein equations and the 
equations of motion for the matter field be satisfied for the $O(1)$ and 
$O(r^2)$ terms separately. See~\cite{mhio2002b,mhio2003} for the 
complete set of the ordinary differential
equations. Both from EOS1 and EOS2 with equations (\ref{zeromu}) and 
(\ref{zerop}), $P_1=W_1=0$ is concluded such that
\begin{eqnarray}
P_1=W_1=0,\quad P_2 =K(8 \pi G)^{1-\gamma}W_2^\gamma, \quad \mbox{(A)}\label{casea2}
\end{eqnarray}
for EOS1, while 
\begin{eqnarray}
P_1=W_1=0,\quad P_2 =\frac{K}{(8\pi G)^{\gamma-1} m_b^{\gamma}}\left(W_
2-\frac{P_2}{\gamma-1}\right)^\gamma, \quad \mbox{(B)}
\label{caseb2}
\end{eqnarray}
for EOS2. From equations (\ref{basic3}), (\ref{basic4}) and 
(\ref{basic6}), we obtain
\begin{eqnarray}
e^{2\Psi}&=&(S+S')^2,\label{eqzero1}\\
3SS^{'2} e^{-2\Phi}&=&P_2S^2S'+W_2S^2(S+S'),\label{eqzero2}
\end{eqnarray}
where we have omitted the bars of ${\bar \Phi}$ and ${\bar \Psi}$ in (\ref{finitessform}) for simplicity and the prime denotes the derivative with respect to $\ln \xi$. 
Equation (\ref{11}) gives
\begin{eqnarray}
2S(S''-\Phi'S')+S^{'2}&=&-P_2S^2 e^{2\Phi},\label{zero11a}\\
(S+S')(S+S'+2\Phi'S)&=&e^{2\Psi}.\label{zero11b}
\end{eqnarray}
$\exp(\Phi)=c_0$ is concluded from equations (\ref{zero11b}) and (\ref{eqzero1}), 
where $c_0$ is a positive constant. Then $P_2=p_0$ is obtained from 
equation (\ref{basic1}), where $p_0$ is a constant, 
which implies that $W_2=w_0$, 
where $w_0$ is a constant. Then, equation (\ref{eqzero2}) gives the 
evolution equation for $S$:
\begin{eqnarray}
\frac{3}{c_0^2}\left(\frac{S'}{S}\right)^2-(p_0+w_0)\frac{S'}{S}-w_0=0. 
\label{evols2}
\end{eqnarray}
The solution to this equation is $S=s_0 \xi^q$, where $s_0$ and $q$ are 
constants. $q \ne -1$ must be satisfied because of equation (\ref{eqzero1}). Equation (\ref{basic2}) with the fact $P_2+W_2 \ne 0$ gives
\begin{eqnarray}
\Psi'S+2S'=0.\label{zero10}
\end{eqnarray}
Then, the equality $q=0$ can be obtained from equation (\ref{eqzero1}) 
and (\ref{zero10}), which implies that $S=s_0$. Finally, equations (\ref{zero11a}) and (\ref{evols2}) give $p_0=0$ and $w_0=0$, respectively. 
Therefore, it is concluded that cases (A) and (B) result in vacuum 
spacetimes.  

\subsubsection{Infinite kind}
Finally, we consider the infinite-kind kinematic self-similar solutions 
in the tilted case. In this case, the Einstein equations imply that the 
quantities $\mu, p$ and $m$ are of the forms
\begin{eqnarray}
2Gm&=&M_1(\xi)/t^2+M_2(\xi),\label{infm}\\
8\pi G \mu&=&W_1(\xi)/t^2+W_2(\xi),\label{infmu}\\
8\pi G p&=&P_1(\xi)/t^2+P_2(\xi),\label{infp}
\end{eqnarray}
where $\xi=r/t$. A set of ordinary differential equations is obtained 
when it is demanded that the Einstein equations and the equations of 
motion for the matter field be satisfied for the $O(1)$ and $O(t^{-2})$ 
terms separately. See~\cite{mhio2002b,mhio2003} for the complete 
set of ordinary differential equations. 
Both from EOS1 and EOS2 with equations (\ref{infmu}) and (\ref{infp}), $
P_1=W_1=0$ is concluded such that
\begin{eqnarray}
P_1=W_1=0,\quad P_2 =K(8\pi G)^{1-\gamma}W_2^\gamma, \quad \mbox{(A)}
\end{eqnarray}
for EOS1, while 
\begin{eqnarray}
P_1=W_1=0,\quad P_2 =\frac{K}{(8\pi G)^{\gamma-1} m_b^{\gamma}}\left(W_
2-\frac{P_2}{\gamma-1}\right)^\gamma, \quad \mbox{(B)}
\end{eqnarray}
for EOS2. From equations (\ref{basic5}), (\ref{00}) and (\ref{11}), we 
obtain 
\begin{eqnarray}
S''&=&S'(\Phi'+\Psi')\label{inf11},\\
(1-W_2S^2)e^{2\Psi}&=&2SS''+S^{'2}-2\Psi'S'S,\label{inf00b}\\
(1+P_2S^2)e^{2\Psi}&=&S'(2\Phi'S+S'),\label{inf11a}
\end{eqnarray}
where we have omitted the bars of ${\bar \Phi}$ and ${\bar \Psi}$ in (\ref{infinitessform}) for simplicity the prime denotes the derivative with respect to $\ln \xi$. From 
equations (\ref{inf11}), (\ref{inf00b}) and (\ref{inf11a}), 
\begin{eqnarray}
P_2+W_2=0,\label{infkey2}
\end{eqnarray}
is obtained, which implies that $p=-\mu$ and gives a contradiction. 
Therefore, it is concluded that cases (A) and (B) result in vacuum 
spacetimes.  

As a result, it is shown that there is no kinematic self-similar 
solutions with a nontrivial polytropic equation of state in the tilted 
case. 
\subsection{Nontilted cases}
Even when the parallel and orthogonal cases are considered, except 
for the infinite-kind kinematic self-similar solutions in the parallel 
case which include all static solutions, the 
only possible solutions are the flat FRW solution as a zeroth-kind 
kinematic self-similar solution in the orthogonal case both for EOS1 and 
EOS2 and the closed FRW solution as a second-kind kinematic self-similar 
solution with an index $\alpha=3/2$ in the orthogonal case for EOS2 with 
$\gamma=1/\alpha=2/3$.

\setcounter{equation}{0}
\section{Summary}\label{sec4}
\setcounter{equation}{0}
Self-similarity has been applied to many aspects of physics and 
other scientific fields.
The introduction of self-similarity into Newtonian gravity 
is straightforward because it 
postulates absolute space and time.
Since Newtonian gravity has only one 
dimensional constant, i.e. the gravitational constant, 
we can incorporate a polytropic gas as well as
an isothermal gas into the framework of complete similarity.
The introduction of self-similarity into general relativity 
is, however, not so straightforward because there is no preferred 
coordinate system in the theory.
The covariant definition of complete similarity 
in general relativity is homothety.
Moreover, since two dimensional physical constants,
the gravitational constant and the speed of light, 
are included in general relativity, 
it is impossible to incorporate many physically interesting matter 
fields, such as a polytropic equation of state,
into the framework of homothety.
This naturally leads to the introduction of incomplete similarity 
in general relativity. One of the most natural definitions 
of incomplete similarity in the fluid system in general relativity
is kinematic self-similarity.
Many known exact solutions turn out to be kinematic self-similar.
At first glance it seems possible to construct kinematic 
self-similar solutions with a polytropic equation of state. 
However, more comprehensive study of the Einstein equations 
reveals that there are no such solutions.
Although the present discussion implies
somewhat limited application of kinematic self-similarity,
there still remains a large possibility that 
kinematic self-similar solutions 
describe interesting gravitational phenomena of physically important
matter fields, such as a double fluid system~\cite{coleyprivate}.


\section*{Acknowledgment}
We would like to thank H.~Iguchi and N.~Okuyama for stimulating discussions. 
TH is grateful to A.A.~Coley for helpful discussion.
This work was partially supported by a Grant for The 21st Century COE Program (Holistic Research and Education Center for Physics 
Self-Organization Systems) at Waseda University.
TH was supported by JSPS Fellowship for Research Abroad.


\begin{table}[htbp]
	\caption{The complete list of kinematic 
self-similar solutions with a perfect fluid in the spherically symmetric
 spacetime for a perfect fluid with EOS1, EOS2 and EOS3. 
It is assumed that the energy density of the perfect fluid is not negative.
See text and references therein.}
	\label{tablesol4}
	\begin{center}
		\begin{tabular}{|l|l|l|}
		\hline \hline
		Matter field & Kind & Solution  \\ \hline\hline
		Vacuum & first, parallel & Minkowski \\ 
		& first, orthogonal & Minkowski  \\ 
		& second, tilted, any $\alpha$ & Minkowski\\
		& second, tilted, $\alpha=3/2$ & Schwarzschild \\ 
		& zeroth, tilted & Minkowski  \\ 
		& zeroth, orthogonal & Minkowski  \\ 
		& infinite, parallel & Minkowski  \\
		&              & Schwarzschild  \\ \hline
Cosmological constant       & zeroth, tilted & de Sitter \\ 
		& zeroth, parallel & de Sitter \\ 
		& zeroth, orthogonal & de Sitter \\ 
		& infinite, tilted & Nariai \\ 
		& infinite, parallel & de Sitter \\ 
		&                    & Nariai \\ 
		&                    & Schwarzschild-de Sitter \\ \hline
		Dust & first, tilted & see Ref.~\cite{carr2000} \\ 
		& second, tilted & KSS LTB \\ 
		& second, parallel, $\alpha=3/2$ & Flat FRW \\ 
		& zeroth, tilted & KSS LTB \\ 
		& zeroth, orthogonal & Flat FRW \\ 
		& infinite, tilted & KSS LTB \\ \hline
		Perfect fluid: EOS1 & zeroth, orthogonal & Flat FRW  \\ 
		& infinite, parallel & All static solutions \\ \hline
		Perfect fluid: EOS2 & second, parallel, $\alpha=3/2$ & Closed FRW
		 with $\gamma=2/3$ \\
		& zeroth, orthogonal & Flat FRW  \\ 
		& infinite, parallel & All static solutions \\ \hline
		Perfect fluid: EOS3 & first, tilted & see Refs.~\cite{op1990,cc2000} \\
		& first, parallel & FRW ($K=-1/3$) \\  
		& first, orthogonal & Gutman-Bespal'ko ($K=1$) \\  
		& second, tilted, $\alpha \ne 3(1+K)/2$ & Flat FRW \\
		& second, tilted, $\alpha \ne 2K/(1+K)$ & Homothetic static\\
		& second, parallel, $\alpha=3(1+K)/2$ & Flat FRW \\ 
		& second, orthogonal, $\alpha=2K/(1+K)$ & Homothetic static \\
		& zeroth, orthogonal, & Flat FRW \\ 
		& infinite, tilted, & Flat FRW \\ 
		& infinite, parallel & All static solutions\\ \hline\hline
		\end{tabular}
	\end{center}
\end{table}

\newpage
\addcontentsline{toc}{section}{\protect\numberline{}{References}}

\end{document}